\documentclass[final,3p,times,twocolumn]{elsarticle}

\usepackage{blindtext, graphicx, amsmath, algorithm, algpseudocode, pifont, algcompatible, comment, layout, amsthm, amssymb}
\usepackage{enumitem}   
\usepackage{eso-pic}
\usepackage{booktabs}
\usepackage{float}
\usepackage{subfigure}
\usepackage{titlesec}
\titlespacing{\section}{2pt}{*1}{*1}
\titlespacing{\subsection}{0pt}{*2}{*0}

\usepackage[utf8]{inputenc}
\usepackage[english]{babel}
\usepackage{hyperref} 
\usepackage{xcolor, soul}
\sethlcolor{green}
\hypersetup{ colorlinks=true, linkcolor=black, filecolor=black, urlcolor=cyan, }
\usepackage{caption}
\usepackage{xcolor}

\journal{ICT Express}

\begin{document}

\begin{frontmatter}

\title{Massive MIMO NOMA with wavelet pulse shaping to minimize undesired channel interference}
\author{Muneeb Ahmad}
\ead{muneeb.ahmad@kumoh.ac.kr}

\author{Soo Young Shin \corref{cor1}}
\ead{wdragon@kumoh.ac.kr}

\address{Department of IT Convergence Engineering, Kumoh National Institute of Technology, Gumi, South Korea}

\cortext[cor1]{Corresponding author}

\begin{abstract}
In this article, wavelet Orthogonal Frequency Division Multiplexing (OFDM) based non-orthogonal-multiple-access (NOMA) combined with massive MIMO system for 6G network is proposed. For mMIMO transmissions, the suggested system could enhance the performance by utilizing wavelets and is able to compensate channel impairments for the transmitted signal. Performance measures include spectral efficiency (SE), symbol error rate (SER), and peak to average power ratio (PAPR). Simulation results prove that the proposed system outperforms the conventional fast Fourier transform (FFT) based NOMA systems.
\end{abstract}

\begin{keyword}
NOMA \sep 6G \sep mMIMO \sep Imperfect SIC \sep Wavelet filter banks \sep FFT-OFDM \sep Future Wireless Communication

\end{keyword}

\end{frontmatter}


\section{Introduction}\label{sec1}
Future wireless networks, such as sixth-generation (6G), is projected to enable exceptionally high data rates and a huge number of users with a wide range of applications and services \cite{1}. Massive multiple-input-multiple-output (mMIMO) is considered as one of the most suitable technology for 6G because it provides higher spectral efficiency via spatial diversity and by allowing its antenna array to focus narrow beams towards a user. Specifically, by deploying a large number of antennas and utilizing the space domain to multiplex various users, the mMIMO technology has the ability to significantly reduce system latency, and to deliver exceptional connection improvements \cite{2}. Similarly, the millimeter-wave (mmWave) transmission is considered as another key technology to deliver multi-gigabit-per-second transmission throughput and large data capacity \cite{3}. The availability of a large amount of unlicensed bandwidth that allows for Gigabit (Gb) data rate transmission is the main attraction to utilize mmWave communication systems. The mmWave frequencies extend from 30 GHz to 300 GHz, where the Federal Communications Commission (FCC) allocated the spectrum of 57–64 GHz, with a carrier frequency of 60 GHz and a bandwidth of 7 GHz for mmWave communication. Various key enabling technologies for mmWave communication are presented in \cite{3}.
\par
Non-orthogonal-multiple-access (NOMA) is also proposed as a viable choice for 5G wireless communication networks, in which the transmit power is exploited to segregate the signals of different users \cite{4,5}. The path loss difference amongst users is utilized to allocate power, where no additional processing is required at the receiver. The user data is multiplexed using superposition algorithm and transmitted through the channel from the base station (BS). At the receiver side, a successive-interference-cancellation (SIC) technique is used to recover the user data. The high power signal is retrieved and subtracted from the received signal that leaves the less power user only to recover and decode its data. While, the high power user directly recover and decodes its own data by considering less power users' signal as noise \cite{5}. 
\par
NOMA is proposed as multi-user superposition transmission (MUST) for long term evolution advanced (LTE-A) networks (3GPP version 13), and according to the LTE-A Pro (3GPP release 14), the standardization of NOMA is considered for uplink side, particularly in massive machine type communication (mMTC) \cite{6}. In addition, certain link-level and system-level performance analysis that establish the feasibility of deploying NOMA schemes, as well as thorough studies for capacity and sum-rate enhancement for mMIMO-NOMA relay systems, have been presented in the literature \cite{2,7,8}.
\begin{figure*}[t!]
\centering
\includegraphics[width=12cm,height=8cm]{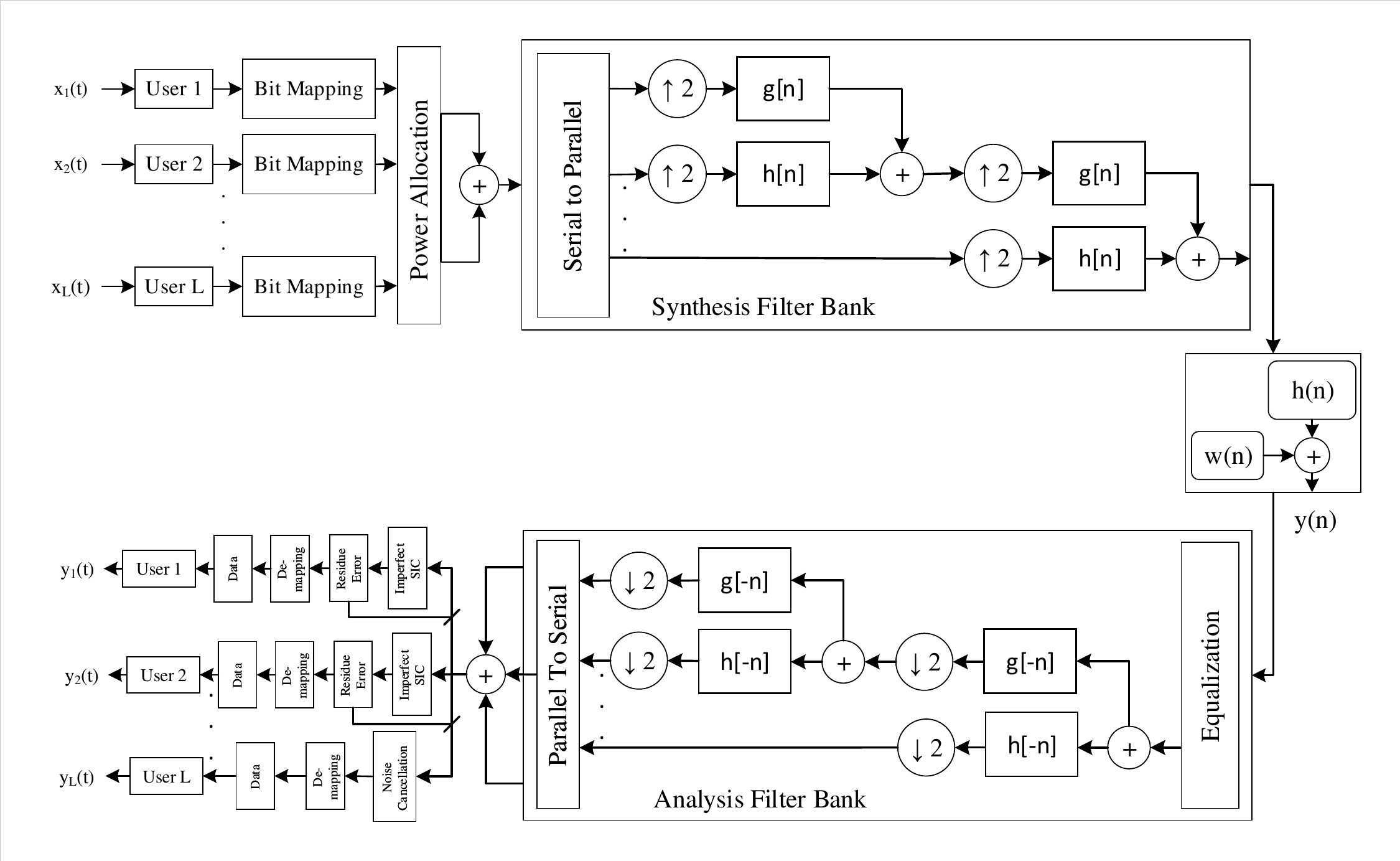}
\caption{The Transceiver Structure for multi-user Wavelet based NOMA system: (Residual Error Approach)}
\label{fig:mMIMO-NOMA}
\end{figure*}
\subsection{Prior Works}
NOMA and mMIMO are integrated to pave the way for the development of 5G/B5G by enhancing the spectral efficiency (SE), and the energy efficiency (EE) of the cellular networks \cite{9}. The feasibility of NOMA and its performance with the mMIMO setup is first mentioned in \cite{10}, where the comparison of conventional mMIMO setup and NOMA is performed in Rayleigh fading channel. Likewise, the NOMA architecture in a multiple-input-single-output (MISO) downlink scenario is studied in \cite{11}, where, the effect of quasi-degradation on NOMA downlink transmission is investigated. To increase overall system performance, a low-complexity sequential user pairing method is devised by using the characteristics of hybrid NOMA precoding. Similarly, \cite{12} presents a downlink beamforming (BF) for hybrid NOMA networks to counteract inter and intra-cluster interference.
\par
Fast Fourier transform (FFT) based OFDM multiple access scheme is implemented in the traditional MIMO or mMIMO-NOMA systems, where the exponentially rising data-rate demands have limited the spectrum utilization in FFT-OFDM based networks. A substantial drawback of FFT-OFDM technique is the limited number of user connectivity to the network due to the limited orthogonal resource allocation and scheduling \cite{13}. Moreover, the sinc function properties of the FFT filter banks makes it more prone to the undesired energy in the side lobes that spills over into the adjacent sub-carriers. In addition, the cyclic prefix (CP) is added to counter the inter user interference and other channel impairments, and this redundancy of CP bits leads to the loss of overall system's throughput and spectral efficiency. The literature on wavelet OFDM-based pulse shaping for NOMA, on the other hand, provides either lower SER or low latency with improved spectral efficiency \cite{14,15}.
\subsection{Motivation and Contribution}
The aforementioned literature study assumed that the number of transmit antennas was either equal to or less than the number of receiving antennas. The appropriate analysis of a mMIMO-NOMA system with a large number of transmit antennas was left unaddressed. Moreover, the mMIMO-NOMA setup was not properly analyzed and compared for different pulse shaping techniques to validate the performance comparison of the system. The standard FFT-based pulse shaping is assumed in the literature and it does not appear to be well suited for mMIMO networks because of the CP insertion and high PAPR, which causes redundant bandwidth utilization and makes it more sensitive to the multipath fading effect. As a result, it may not be the best recourse for future 6G wireless communication networks, which demand increased spectral efficiency and a huge data rate with reduced SER. Similarly, the SIC for NOMA is always assumed to be perfect in previous studies and the effect of imperfect SIC for mMIMO-NOMA setup is not examined before. To the best of our knowledge, the wavelet-based mMIMO-NOMA (mMIMO-WNOMA) system with imperfect SIC for different pulse shaping techniques is not considered before. 

Therefore, the contribution of this includes:
\begin{itemize}
\item A mMIMO-WNOMA system with large number of transmitting antennas at the BS, which can improve the performance of future 6G wireless communication networks not only in terms of lower SER but also in terms of increased capacity and reduced sensitivity to noise. 
\item Furthermore, the concept of wavelet function and its benefits are provided to the multiuser mMIMO-NOMA system to enhance the SER and PAPR. Since, wavelet-based NOMA has never been examined in a massive MIMO NOMA setup. This design is expected to ensure robustness and spectrum efficiency, as well as interoperability and compatibility with the existing wireless communication systems.
\item Moreover, the presented system is not only analyzed with perfect SIC assumptions but also for the real practical scenario under imperfect SIC in the presence of undesired data due to the incomplete subtraction of the interference at the intended SIC performing user, as shown in Figure 1.
\end{itemize}
\begin{figure}[t!]
\centering
\includegraphics[width=7cm,height=7cm]{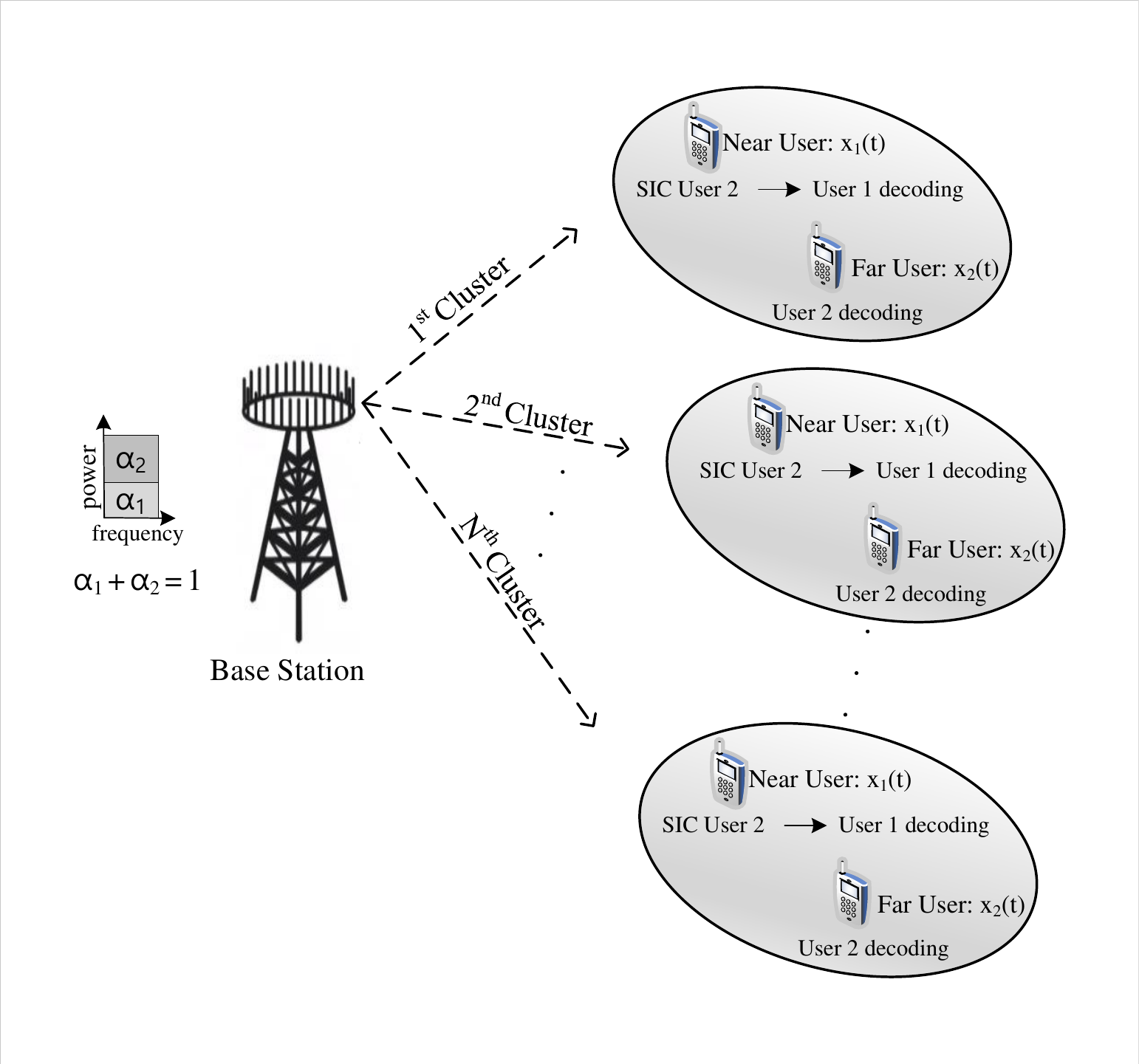}
\caption{System Model for mMIMO-WNOMA}
\label{fig:mMIMO-NOMA}
\end{figure}
\section{System Model for mMIMO-WNOMA}
Consider a downlink mMIMO-NOMA system with multiple antenna at the BS transmitting to the multiple single antenna users. In the presented system, the total number of users (L) are split into clusters (N). The users are kept in an even number for NOMA pairing and are divided into two categories of Near user ($x_{1}$) and Far user ($x_{2}$). Multiple antenna at the BS are able to direct the beam towards each cluster, as shown in the Figure 2. The indices $N_{near}$$\epsilon$$N$ ($N_{near}$ = \{$1,2,...,L/2$\}), $N_{far}$$\epsilon$$N$ ($N_{far}$ = \{$L/2+1,...,L$\}), and ($N$ = \{$1,2,3,...,L$\}) are utilized to represent near and far user in each cluster. The classification of users is required for NOMA pairing due to the channel fading, and it is specifically associated to the large-scale fading effect experienced by the users. For a real practical scenario, in a mMIMO-NOMA system, the near and far users are not completely dependent on the distances from the BS; rather, a far user may experience higher large-scale fading due to shadowing than a near user and may belong to the group index $N_{near}$. The time-division-duplex (TDD) mode is considered for the presented system because of the channel reciprocity (CR). Utilizing the concept of CR, the BS can estimate the downlink channels subjected to the uplink pilots. Hence, the BS is able to perform the beam-forming based on the channel estimation. Zero-forcing (ZF) beam-forming (BF) is considered in this article, because BF is a general approach to utilize for MIMO network with multiple antenna at the BS. Following the system configuration described above, the BS will transmit the beam-formed superimposed signal $x_n$ of $n_{th}$ user in a cluster as:
\begin{equation}
\centering
x_{n}=v_{n} \sum_{i=1}^{2} \sqrt{\alpha_{n, i}} s_{n, i}
\end{equation}
where, $\mathbf{\alpha}_{n,i}$ and $s_{n, i}$ are the allocated power and user data to the $i_{th}$ user in the $N^{th}$ cluster, respectively. $\boldsymbol{v}_{n}$ is the precoding vector for the desired cluster. NOMA users in the same cluster share the same precoding vector but the transmit power is different. The precoding vector for the BF is selected to support multiple users in NOMA, as well as to support the SIC process at the near user, because it is susceptible to interference from neighboring clusters. That is why, the BS needs to know the CSI of the near users from all the clusters. However, this type of ZF-BF implementation does not provide array gain to the far user, rather it supports the SIC process and signal recovery at the near and far user, respectively. 

\section{Conventional NOMA and Wavelet NOMA: A Pulse Shaping Technique}
OFDM technique offers various features i.e. lower complexity, less multi-path delay spread, simple channel equalization techniques and interoperability to the existing MIMO networks. For the real environment, the SIC process is not always perfect due to the unknown channel conditions at the BS. That is why, the BS is unable to provide the complete channel information about the second group of the users ($N_{far}$) to the SIC performing group of users ($N_{near}$) in each cluster. In this situation, the residual error from the initial SIC performing user would spread across the system, degrading the entire signal decoding process.

\par
A general transceiver structure for multi-user wavelet NOMA (WNOMA) is presented in Figure 1, with imperfect SIC effect flowing from the first SIC performing user to the others. After assigning the fractional power to the user's data, the data symbols are subjected to the source coding and modulation. Wavelet pulse shaping is applied at the transmitter side. The wavelet filter bank modulation process at the transmitter begins with the serial to parallel conversion of the data stream into parallel data sub-streams, which are then processed through an octave filter bank. After up sampling the input data, the synthesis filter bank is implemented on the transmitter side. 
The channel is then used to transmit the modulated data. As indicated in Figure 1, the data recovery process comprises of the reverse technique used on the transmitter side. 

Multi-carrier modulation in both FFT-OFDM and wavelet-OFDM based NOMA allows multiple low SER data streams over the sub-carriers. Each beam is assigned by a pre-coding vector and directed towards the desired cluster. For NOMA, the near users in each cluster will perform the SIC after passing the signal through both the equalization and analysis filter bank. While, the far users will directly decode the data by treating the near user's interference as noise. The DFT filter banks are applied to the data for FFT-OFDM based NOMA. Because this article analyzes both Fourier and wavelet transformations in the presence of interference due to imperfect SIC and unknown channel conditions, the following subsections will briefly discuss the reduced interference effect, mitigation of residual error and enhanced PAPR characteristics of wavelet-based NOMA compared to the FFT-based NOMA under mMIMO setup.
\subsection{Reduced Interference affect for mMIMO-WNOMA}
Compared to the FFT-OFDM, wavelets utilize shorter waveform for orthogonal base and that makes the wavelets robust to multipath delay spread those results in inter-symbol-interference (ISI) and inter-carrier-interference (ICI). ISI occurs when the delayed waves of sub-channel deteriorate the reception of the currently transmitted symbol of the same sub-channel. Whereas ICI occurs due to the orthogonality disturbance of the sub-channels. Thus, both the interference affects the SIC process due to the mentioned delays in the sub-carrier and the situation even gets worse for the increased sub-carriers. However, the inherent nature of the wavelets of less sensitivity to offsets in both time and frequency domain plays a vital role to reduce these delays. Hence, the delayed symbols from the other sub-channels are successfully clipped out to eliminate the effect of ICI. Moreover, the customize-able characteristics of the wavelets give an extra benefit to adapt the channel to reduce ISI. The filter response of both the FFT and wavelets can be seen in Figure. 3(d), where lower power requirements for frequency offsets from interferences and a lower PAPR make wavelet-based system model more resistant to interference from imperfect SIC and aid in improved signal reconstruction under mMIMO setup, therefore boosting the system's SER.
\par
\subsection{Residual error and mMIMO-WNOMA}
Traditional NOMA pulse shaping is based on rectangular windows of identical size to the Fourier Transform. While, the wavelet filters in the proposed system utilize inverse discrete wavelet transform (IDWT) and discrete wavelet transform (DWT) at the transmitter and receiver side, respectively. At the transmitter, the encoded symbols are transformed into the wavelet symbols when passed from synthesis filter banks. The input signal is decomposed into low and high pass components $g[n]$ and $h[n]$, respectively as shown in Figure 1. Later the IDWT process is performed and the signal can be represented as \cite{14}:
\begin{equation}
\begin{aligned}
x_{i}=&\sqrt{2^{-(J-1)} \frac{E}{T}} \sum_{j \in \mathrm{I}} a_{j}^{0} \phi\left(2^{-(J-1)}\frac{t}{T}-j \right)\\
&+\sum_{n=1}^{J-1} \sqrt{2^{-(J-n)} \frac{E}{T}} \sum_{j \in \mathrm{I}} a_{j}^{n} \psi\left(2^{-(J-n)} \frac{t}{T}-j\right)
\end{aligned}
\end{equation}
where, $E$ and $T$ are the average symbol energy and the symbol duration respectively. $a^{n}_{j}$ is the complex valued M-ary modulated symbol with $n=0,...,j-1$. $I$ is an integer to represent the index set. $j$ is another integer with positive value to show the multi-level decomposition of the wavelets family. Whereas, $\phi(m)$ and $\psi(m)$ is the scaling and wavelet function respectively. Similarly, $y_{n}$ is the composite signal propagated over the Rayleigh-Fading channel and received at the $n_{th}$ user in a cluster, and can be shown as ($n=1,2$):
\begin{equation}
y_{n}=h_{n}\sum_{i=1}^{2}(s_{i}\sqrt p_{i}) +z_{n},
\end{equation}
where, $h_{n}$ represents the channel coefficient between BS and $i_{th}$ user. $s_i$ and $p_{i}$ are the $i_{th}$ users' intended signal and power, respectively. The channel here is considered to be the product of path loss and Rayleigh fading, with a mean of zero and a variance of one $\operatorname{CN}\left(\mathbf{0}, \mathbf{1}\right)$. Each user's channel vector contains both the large-scale fading coefficient and the small-scale fading vector. Where, in each coherence interval, the large scale fading vector uses one independent Rayleigh fading realization. $z_n$ is the $i_{th}$ channel link's zero-mean complex additive Gaussian noise with $\sigma^2$ as variance. At the receiver end, least square (LS) and minimum-mean-square-equalization (MMSE) equalization is considered for the received signal to nullify the channel effect. Later, the signal is passed through the analysis filter bank for wavelet reconstruction with the inverse process applied at the synthesis filter banks. Now the data is subjected to the SIC process. 
\par
It is important to mention here that most of the literature study assumed perfect SIC that is one of the key aspects to realize the NOMA gain. However, perfect SIC for NOMA is difficult to achieve, since complete channel information at the receiver is not possible in real practice. Therefore, in the presented massive MIMO setup, the interference of real practical scenario is considered. As presented in the literature study \cite{16}, for SIC performing users, an imperfect SIC causes undesired interference in the signal. The intuition behind undesired interference is that the users are assigned power based on their distance from the BS, the far user decodes its data by treating the signal from the near user as noise. However, SIC is carried out at the near user. Due of the incomplete CSI, the near user will be unable to entirely remove the data of the far user. As a result of the incomplete cancellation of the unwanted far user's signal, some undesired data will remain in the signal, which is referred to as residual error. This residual error affects the intended user's SER. Moreover, the flow of residual error through the entire massive MIMO setup can degrade the system's SER performance owing to mass connection as presented in the receiver side of Figure 1. Therefore, Wavelet filters are preferred here in the presented mMIMO-WNOMA system to reduce the influence of residual error since wavelets provide higher resilience to signal distortion caused by unwanted energy as detailed in the previous subsection. To accomplish this, the received signal is processed in the analysis filter bank, where it is passed through low and high pass filters with impulse response $g_n$ and $h_n$, respectively. The use of FFT and DWT is a signal's linear transformation procedure. Where, the received signal can be transformed using FFT as follows:
\begin{equation}
\begin{aligned}
&y(t)=\mathrm{f_{FFT}}[x(t)+\omega(t)] \\
&y(t)=X(t)+\omega_{\mathrm{FFT}}(t)
\end{aligned}
\end{equation}
Similarly, by applying DWT to the received signal, $y(t)$ may be represented as:
\begin{equation}
\begin{aligned}
&y(t)=\mathrm{f_{DWT}}[x(t)+\omega(t)] \\
&y(t)=X(t)+\omega_{DWT}(t)
\end{aligned}
\end{equation}
where, $\omega_{t}$ comprises the interference caused by the residual error, and channel noise $z$. $\omega_{FFT}(t)$ and $\omega_{DWT}(t)$ is noise output from FFT and wavelet filter banks, respectively.

\subsection{Spectral Efficiency and PAPR improvement in mMIMO-WNOMA}
In recent network deployments, either 5G or LTE-A, FFT-OFDM pre-appends the CP to counter interference thus wasting $20$ to $25\%$ of the available bandwidth. In the time domain, wavelet OFDM symbols overlap, and their higher side lobe attenuation makes the communication system robust without requiring CP. Consequently, the spectral efficiency of wavelet based mMIMO-NOMA system is much higher than the conventional FFT based mMIMO-NOMA. The SE comparison for both pulse shaping achemes can be denoted as $E_{FFT}$ and $E_{W}$ \cite{17};
\begin{equation}
E_{FFT}=1 /\left(T+T_{CP}\right) F=T /\left(T+T_{CP}\right)<1
\end{equation}
\begin{equation}
E_{{W}}=1 /(T F)=1
\end{equation}
where $F$ means the sub-carrier spacing, $T$ denotes the symbol duration, and $T_{CP}$ denotes the $CP$ duration. SEI's optimal value is $1$. Because of the CP length $T_{CP}$, WNOMA has the highest optimal spectral efficiency than FFT based NOMA.
\par
Because of the above mentioned CP insertion and increased sub-carrier in traditional NOMA system, it possesses the narrow-band signals that are added constructively, hence this increases the instantaneous peak energy of the signal higher than the average signal power. Moreover, the power amplifier in the hardware creates inter-modulation distortions as well \cite{15}. On the contrary, the PAPR of the wavelet filters is much lesser than the FFT-OFDM because of the energy confinement in frequency domain. According to the definition, the PAPR is the ratio of peak and average power of the signal ($\frac{\rho_{\text {peak }}}{\rho_{\text {avg }}}$). Thus the PAPR of the wavelet based NOMA signal is expressed as \cite{13}:
\begin{equation}
\operatorname{PAPR}=\frac{\max _{n}\left\{\left|S_{W}\right|^{2}\right\}}{E\left\{\left|S_{W}\right|^{2}\right\}}
\end{equation}
where $S_{W}$ is the complex transmitted signal vector, the factor $E$ represents the average value and the factor $\max_n$ shows the maximum value of time index among all instances. The PAPR of any multi-carrier modulation is dependent on the pulse-shape \cite{16}. The selection of suitable pulse-shaping technique can improve the PAPR of the system and can be represented as
\begin{equation}
\mathrm{PAPR} \leq Q_{\max }|\gamma(m)|^{2},
\end{equation}
where, $\gamma(m)$ is the scaling function of the filters and the $Q$ is the total sub-carriers used. As a result of the higher PAPR provided by conventional FFT-based mMIMO-NOMA, mass connectivity can significantly reduce network spectral efficiency. The wavelet family in Figure 3(c), on the other hand, allows for greater flexibility in selecting any wavelet type for massive MIMO NOMA networks in order to improve spectral efficiency and data rate.
\begin{figure*}[t!]
     \centering
     \subfigure[]{%
         \includegraphics[height=5.4cm,width=5.4cm]{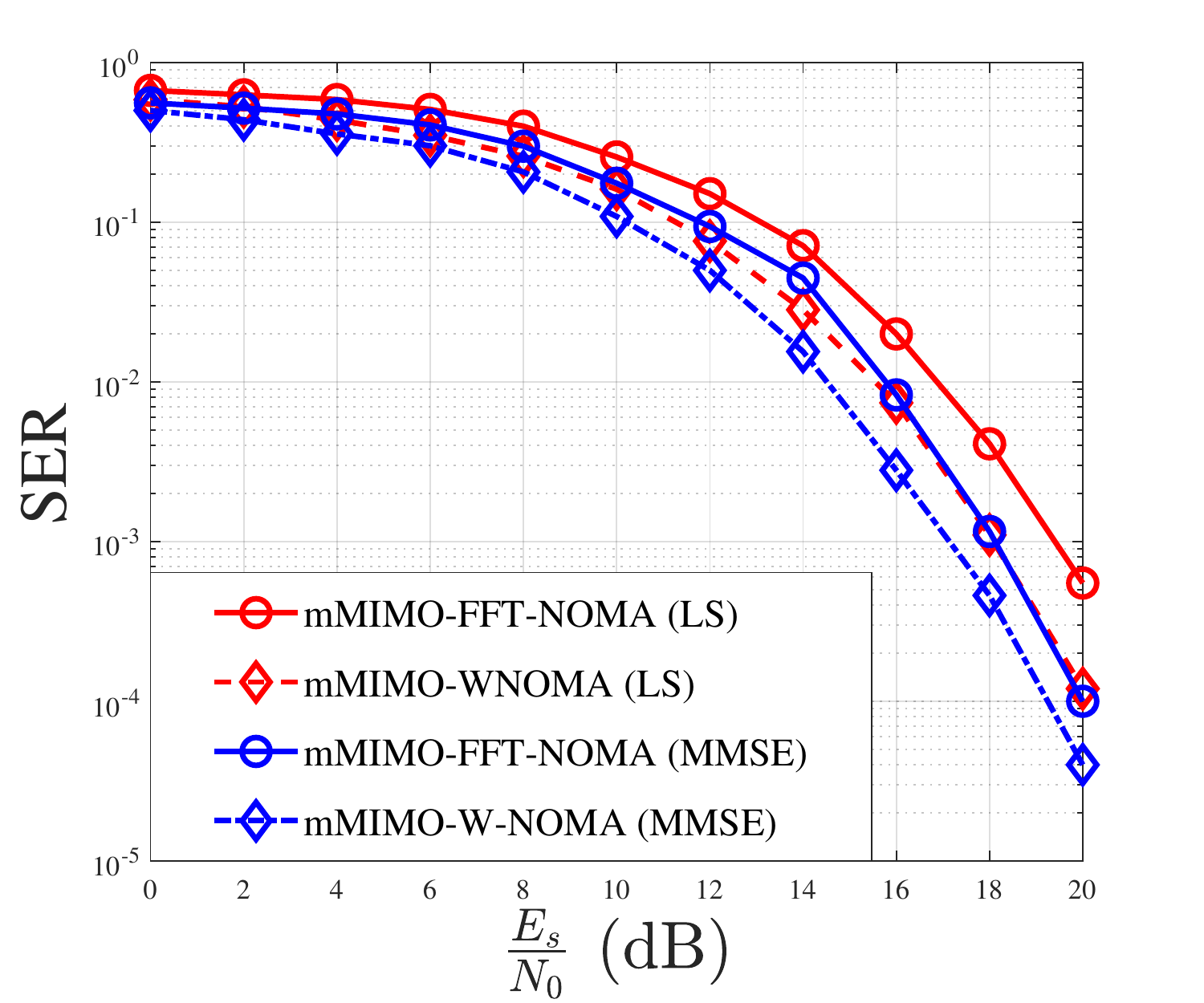}%
         \label{fig:a}%
    }
     \hfill
     \subfigure[]{%
         \includegraphics[height=5.4cm,width=5.4cm]{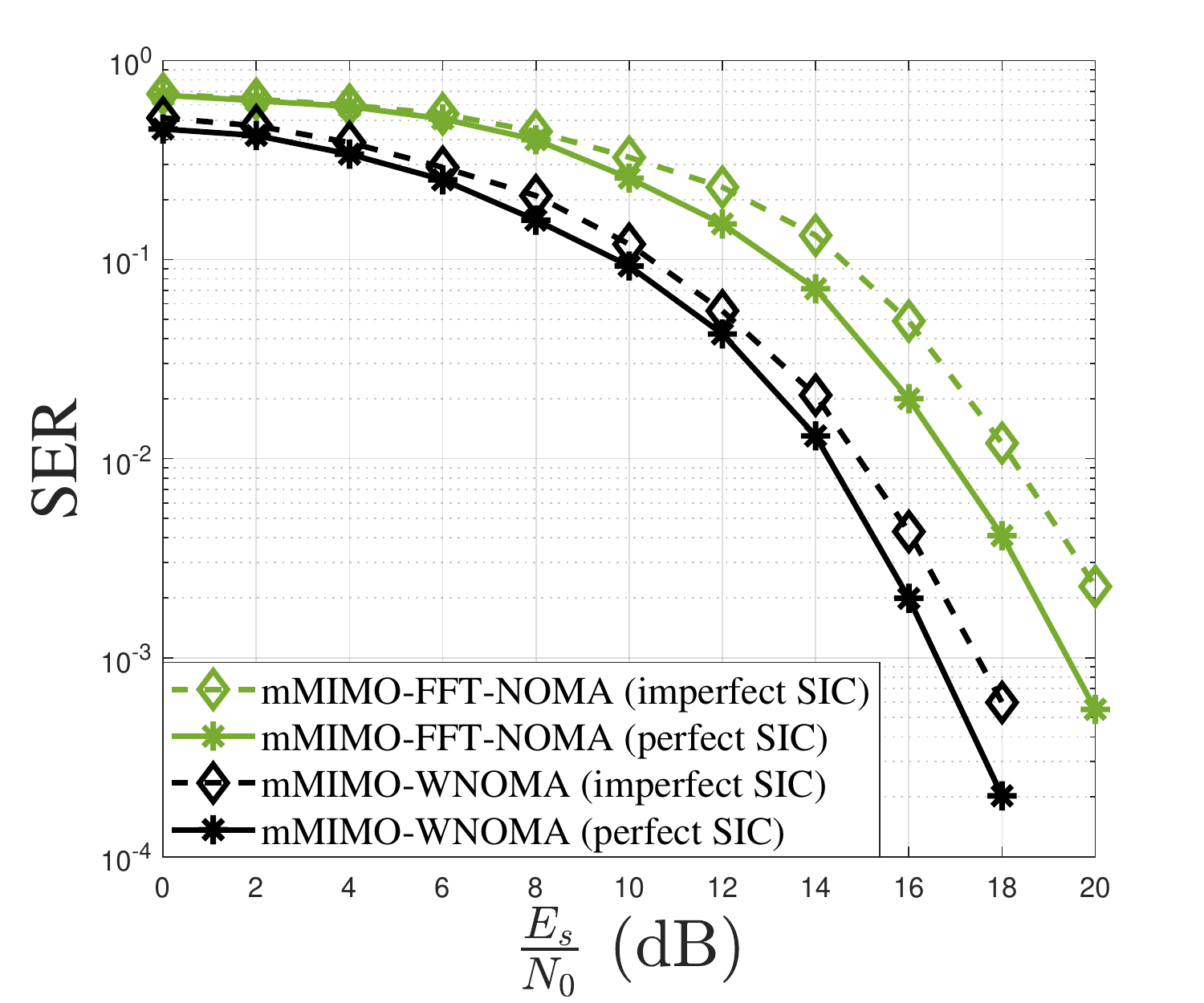}%
         \label{fig:b}%
    }
     \hfill
     \subfigure[]{%
         \includegraphics[height=5.4cm,width=5.4cm]{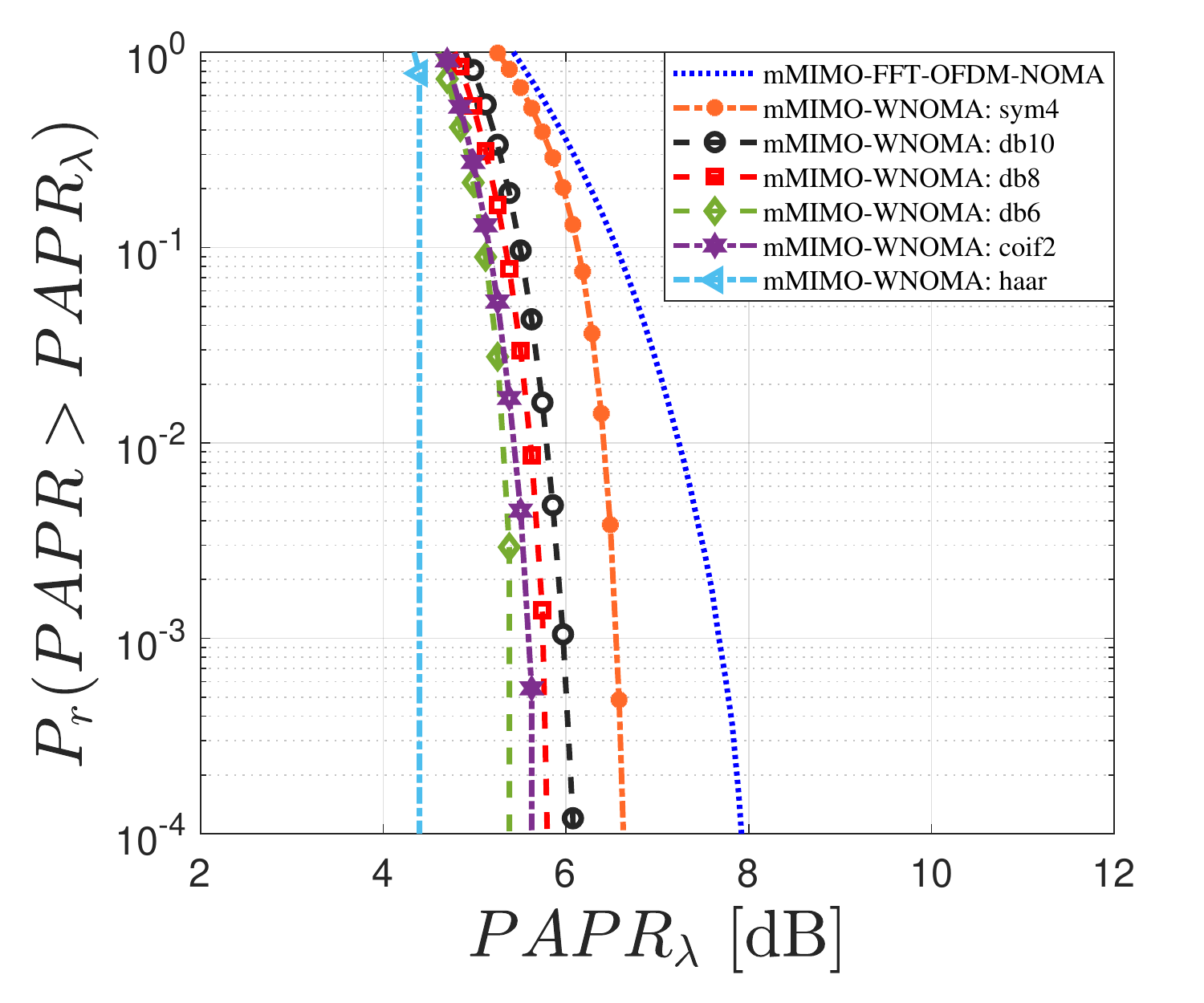}%
         \label{fig:c}%
    }
       \hfill
        \subfigure[]{%
         \includegraphics[height=5.4cm,width=5.4cm]{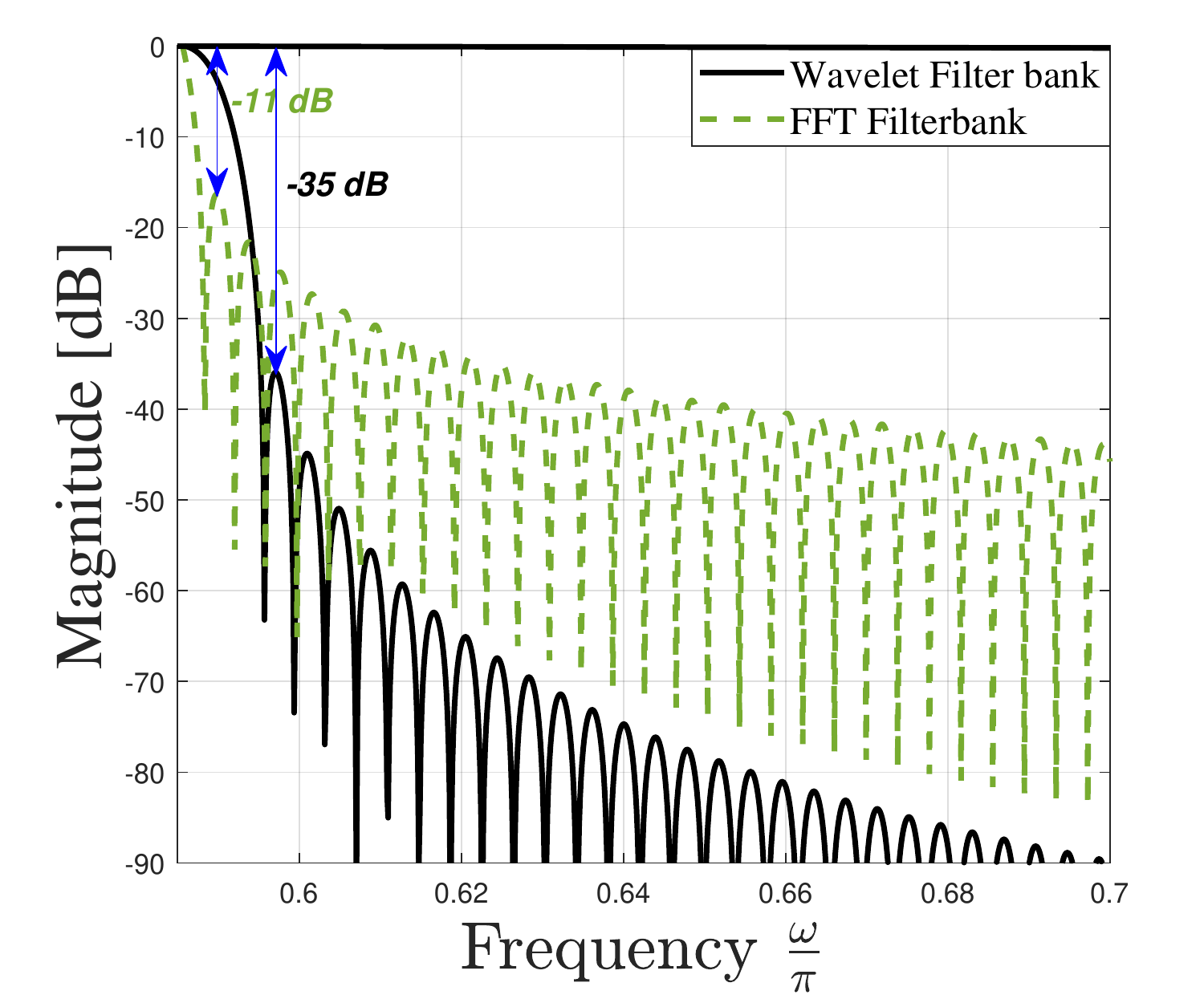}%
         \label{fig:d}%
    }
      \hfill
        \subfigure[]{%
         \includegraphics[height=5.4cm,width=5.4cm]{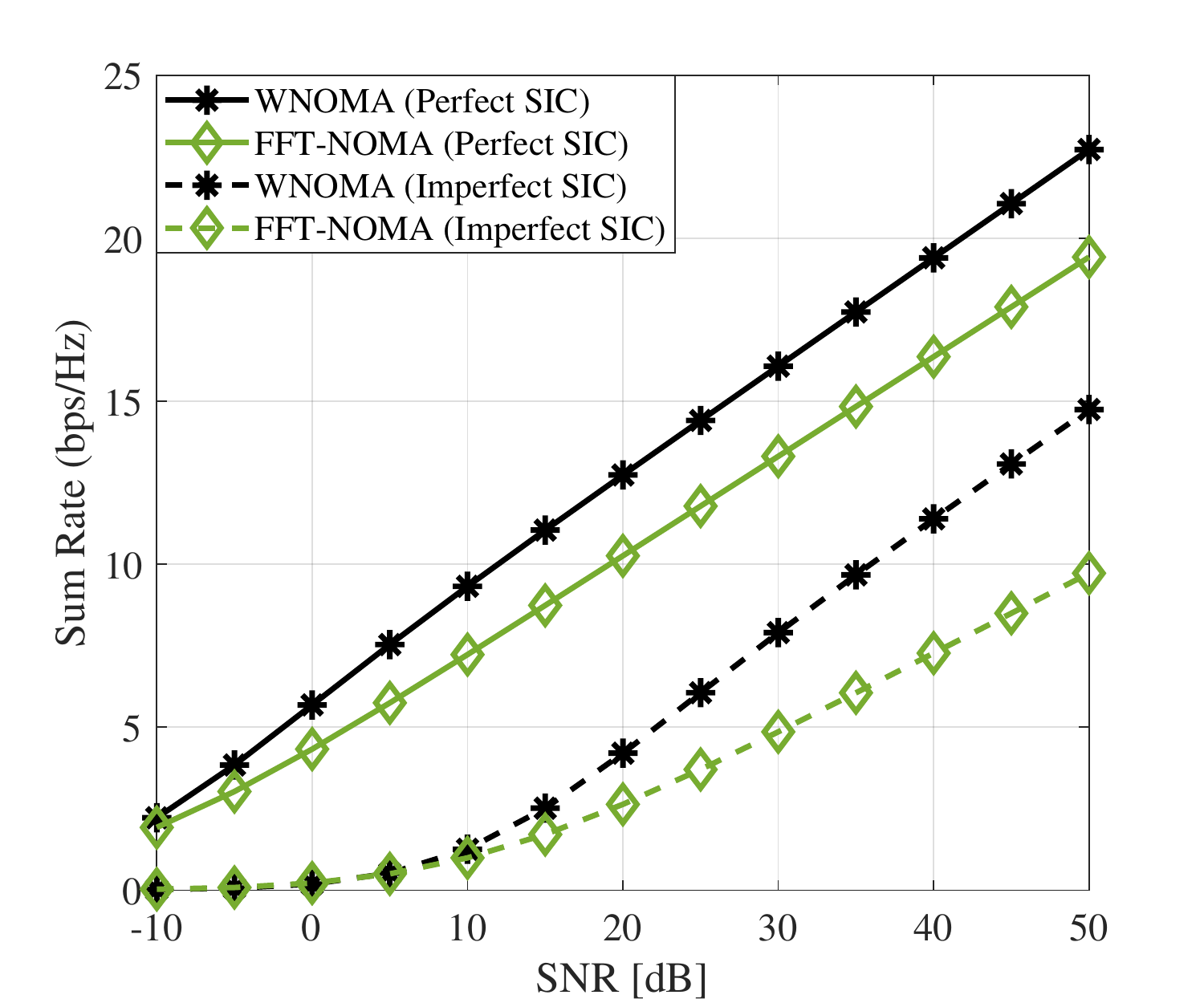}%
         \label{fig:e}%
    }
     \hfill
        \subfigure[]{%
         \includegraphics[height=5.4cm,width=5.4cm]{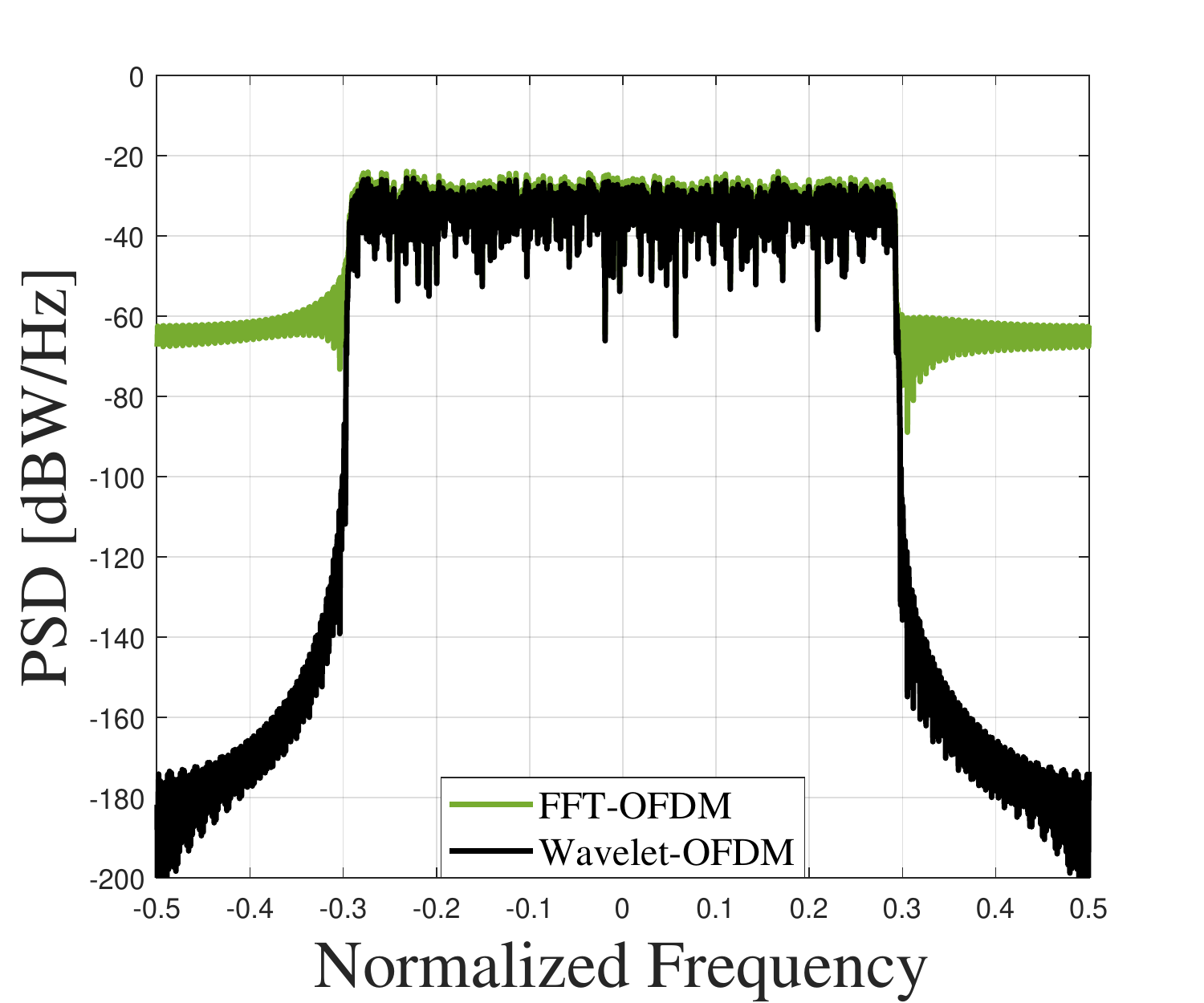}%
         \label{fig:f}%
    }
\caption{SER comparison of FFT-OFDM and wavelet-OFDM based massive MIMO-NOMA (a) LS and MMSE Channel Equalization. (b) Perfect and Imperfect SIC. (c)  PAPR comparison of FFT-OFDM with diverse wavelet family. (d) Magnitude Spectra of FFT and Wavelet filter-banks. (e) Capacity comparison of FFT-NOMA and WNOMA. (f) PSD comparison of FFT-NOMA and WNOMA.}
\end{figure*}
\subsection{MODEM Complexity for mMIMO-WNOMA}
The computational complexity of an FFT-OFDM pulse shaped data is determined by the fast algorithm that computes the Discrete Fourier Transform (DFT). The split-radix FFT algorithm is considered as the fastest DFT algorithm. By following this algorithm, there are $Xlog_2X$ multiplications and $3Xlog_2X$ additions necessary to compute FFT for a complex frequency \cite{15}. On the other hand, the wavelets represent a distinct scenario in terms of computing complexity. The complexity of a wavelet packet transform grows as the number of nodes increases. Since all the nodes of wavelets are traversed in a binary tree of depth $log_2X$, the complexity of wavelets will be $CXlog_2X$ multiplications and additions \cite{14,15}. On that account, the complexity of both FFT and wavelet-based NOMA will be the same. Therefore, owing the same complexity, wavelets offer better PAPR, enhanced SE and reduced SER to meet almost all the requirements of future 6G wireless network.

\section{Simulation Results and Performance Analysis}
The performance of both FFT and wavelet OFDM-based pulse shaping techniques with mMIMO-NOMA setup is evaluated and compared for both perfect and imperfect SIC conditions. The simulation is run in Matlab 2021(a), and the provided mMIMO configuration takes into account 512 transmit antennas at the BS and 1024 user equipment at the receiving end. For wavelet-based pulse shaping, two level decomposition for wavelet filters and 256 sub-channels for FFT filter banks are used. Furthermore, as noted in Section II, each NOMA cluster comprises of two users. As a result, channel gain is set to $-10$ dB for near users and $-5$ dB for far users in these clusters. The $\mathcal{M}-QAM$  ($\mathcal{M}$ = $16$) modulation system is employed for both FFT and wavelet setup, with FFT CP and bandwidth ratios of 20$\%$ and 80$\%$, respectively. It is assumed that the channel conditions are partially known to the receiver for Rayleigh Fading channel with additive white Gaussian noise with reference to the PAPR and SER. 
\par
Figure 3(a) presents the SER curves of traditional NOMA and the proposed WNOMA under mMIMO setup, for the LS and the MMSE technique with perfect SIC conditions. It can be seen that WNOMA outperforms FFT based NOMA in both MMSE and LS channel equalization at the receiver. It is because of the robustness of WNOMA to the interference from neighbouring sub-carriers. Figure 3(b) shows the SER under incomplete CSI scenario, that leads the entire system to the imperfect SIC. As the channel conditions are not completely known at the receiver, the residual error due to the imperfect SIC will lead the entire network to the deteriorated data recovery. It can be seen that the analysis filter banks at the receiver preforms better compared to the most adapted FFT-filter banks for NOMA. WNOMA offers a value of $5.956$x$10^{-4}$ of SER at $20$ dB under imperfect SIC, but the FFT-NOMA delivers its performance of value $2.28$x$10^{-3}$ only. This is because of the lesser PAPR and lower side lobe energy provided by the wavelets as shown in Figure 3(c) and Figure 3(d), respectively. Figure 3(d) verifies that lower power requirements for frequency offsets from interference and lower PAPR make our wavelet-based system model more immune to interference from imperfect SIC and aid in enhanced signal reconstruction, hence improving the system's SER. Wavelet filter bank provides $-24$ dB gain over the FFT-filters and improves spectral efficiency owing to the absence of CP which is the redundancy in FFT-NOMA, and so can maintain high performance under network congestion. 
\par
Figure 3 (c) gives an overview of the PAPR showed by the different wavelets, and it gives an extra degree-of-freedom to choose from the large wavelet family to utilize at the receiver. The symlet-4 and the Haar wavelet has $1.2$ dB and $3.4$ dB gain over the traditional FFT-NOMA. Whereas, the coiflets-2, daubechies-6, 8 and 10 holds the $2.2, 2.3, 2.1$ and $2$ dB gain, respectively. Furthermore, the out-of-band (OOB) radiated noise because of the poor FFT-filter response degrades the SER and spectral efficiency, where the wavelet filter is capable to reduce OOB radiation due to its tight filter taps. Therefore, the analysis of the spectral efficiency in terms of the sum-rate is also given in the Figure 3(e) supported by the power-spectral-density (PSD) plot in Figure 3(f). Figure 3(e) clearly illustrates that WNOMA outperforms FFT-based NOMA in terms of throughput for two different sets of channel circumstances. 
\par
Moreover, Figure 3(f) validates the statement made in subsection 3.3 that the WNOMA gives high spectral confinement as it holds PSD of $-174.19$ dB while the FFT-NOMA holds the value of $-64.4$ dB. When compared to FFT-OFDM, the PSD of wavelet filter banks is quite confined. Because OFDM systems have high side lobes, they have high out of band energy radiation, which results in an increased ICI and ISI. OFDM-based devices pre-append the CP to battle ISI by wasting 20 to 25 percent of the available bandwidth. As a result, the spectral efficiency of OFDM-based NOMA is around $20\%$ worse. Similarly, under imperfect SIC scenario, the WNOMA and the FFT-NOMA provides $14.7$ bps/Hz and $9.7$ bps/Hz of sum-rate value, respectively. Hence, WNOMA offers the added benefit of delivering reliable communication without the need of CP, and allowing the increased multi-user capacity. The presented analysis verifies the enhanced performance of mMIMO-WNOMA for both imperfect and perfect SIC conditions due to its excellent filter response and lesser bandwidth consumption.
\section{Conclusion}
In this article, authors present the wavelet NOMA based massive MIMO system for the future 6G wireless network. The efficient bandwidth utilization can be achieved by implementing NOMA with the mMIMO setup and the affect of unknown channel conditions at the receiver can be minimized via wavelet filter banks. The presented results verify that WNOMA enables the better data recovery compared to the conventional NOMA under both perfect and imperfect SIC conditions. Furthermore, the presented model can improve the capacity of a multi-user mMIMO NOMA system, which can enable high data rates and mass connectivity in a future 6G wireless communication system.
\par
In future works the assumptions can be sought to include more users in a cluster or utilizing the relays for multiple groups in a cluster. Furthermore, as a future extension of this work, the feasibility of implementing a wavelet-based pulse shaping approach to current research on mMIMO-NOMA will be investigated in order to validate the interoperability of the given system and improve the key performance indicators of future cellular networks.
\section*{Acknowledgments}
This work was supported by the National Research Foundation of Korea(NRF) grant funded by the Korea government. (MSIT) (No. 2022R1A2B5B01001994).

\section*{Conflict of interest}
The authors declare that there is no conflict of interest in this paper.



\bibliographystyle{elsarticle-num}



\begin{thebibliography}{gg}

\bibitem{1}
J. Zhu, M. Zhao, S. Zhang and W. Zhou, "Exploring the road to 6G: ABC — foundation for intelligent mobile networks," \textit{China Communications}, vol. 17, no. 6, pp. 51-67, June 2020.

\bibitem{2}
D. Zhang, Y. Liu, Z. Ding, Z. Zhou, A. Nallanathan and T. Sato, "Performance Analysis of Non-Regenerative Massive-MIMO-NOMA Relay Systems for 5G," \textit{IEEE Transactions on Communications}, vol. 65, no. 11, pp. 4777-4790, Nov 2017.

\bibitem{3}
W. Hong et al., "The Role of Millimeter-Wave Technologies in 5G/6G Wireless Communications," \textit{IEEE Journal of Microwaves}, vol. 1, no. 1, pp. 101-122, Jan. 2021.

\bibitem{4}
X. Pei, H. Yu, M. Wen, S. Mumtaz, S. Al Otaibi and M. Guizani, "NOMA-Based Coordinated Direct and Relay Transmission With a Half-Duplex/ Full-Duplex Relay," \textit{IEEE Transactions on Communications}, vol. 68, no. 11, pp. 6750-6760, Nov. 2020.

\bibitem{5}
O. Maraqa, A. S. Rajasekaran, S. Al-Ahmadi, H. Yanikomeroglu and S. M. Sait, "A Survey of Rate-Optimal Power Domain NOMA With Enabling Technologies of Future Wireless Networks," \textit{IEEE Communications Surveys \& Tutorials}, vol. 22, no. 4, pp. 2192-2235, Fourthquarter 2020.

\bibitem{6}
H. Lee, S. Kim and J. Lim, "Multiuser Superposition Transmission (MUST) for LTE-A systems," \textit{IEEE International Conference on Communications (ICC)}, 2016, pp. 1-6.

\bibitem{7}
Khan, A., Usman, M. A., Usman, M. R., Ahmad, M., and Shin, S.Y. "Link and system-level \textit{NOMA} Simulator: The Reproducibility of Research," \textit{Electronics}, vol. 10, no. 19, pp 2388, 2021. 

\bibitem{8}
X. Li et al., "Hardware Impaired Ambient Backscatter NOMA Systems: Reliability and Security," in IEEE Transactions on Communications, vol. 69, no. 4, pp. 2723-2736, April 2021, doi: 10.1109/TCOMM.2021.3050503.

\bibitem{9}
S. Kusaladharma, W. -P. Zhu, W. Ajib and G. A. A. Baduge, "Achievable Rate Characterization of NOMA-Aided Cell-Free Massive MIMO With Imperfect Successive Interference Cancellation," \textit{IEEE Transactions on Communications}, vol. 69, no. 5, pp. 3054-3066, May 2021.

\bibitem{10}
K. Senel, H. V. Cheng, E. Björnson and E. G. Larsson, "NOMA Versus Massive MIMO in Rayleigh Fading," \textit{IEEE 20th International Workshop on Signal Processing Advances in Wireless Communications (SPAWC)}, 2019, pp. 1-5.

\bibitem{11}
Z. Chen, Z. Ding, X. Dai and G. K. Karagiannidis, "On the Application of Quasi-Degradation to MISO-NOMA Downlink," \textit{IEEE Transactions on Signal Processing}, vol. 64, no. 23, pp. 6174-6189, Dec.1, 2016.

\bibitem{12}
Z. Chen, Z. Ding and X. Dai, "Beamforming for Combating Inter-cluster and Intra-cluster Interference in Hybrid NOMA Systems," \textit{IEEE Access}, vol. 4, pp. 4452-4463, 2016.

\bibitem{13}
Khan, A, and Shin, S, Y. “Wavelet OFDM-Based Non-Orthogonal Multiple Access Downlink Transceiver for Future Radio Access.” \textit{IETE Technical Review} vol. 35, no. 1, pp. 17–27, 2016.

\bibitem{14}
S. Baig, U. Ali, H. M. Asif, A. A. Khan and S. Mumtaz, "Closed-Form BER Expression for Fourier and Wavelet Transform-Based Pulse-Shaped Data in Downlink NOMA," \textit{IEEE Communications Letters}, vol. 23, no. 4, pp. 592-595, April 2019.

\bibitem{15}
S. Baig, M. Ahmad, H. M. Asif, M. N. Shehzad and M. H. Jaffery, "Dual PHY Layer for Non-Orthogonal Multiple Access Transceiver in 5G Networks," \textit{IEEE Access}, vol. 6, pp. 3130-3139, 2018.

\bibitem{16}
Ahmad, M., Baig, S., Asif, H. M., and Raahemifar, K. “Mitigation of Imperfect Successive Interference Cancellation and Wavelet-Based Nonorthogonal Multiple Access in the 5G Multiuser Downlink Network.” \textit{Wireless Communications and Mobile Computing},2021, 1–11.

\bibitem{17}
Fa-Long Luo; Charlie Zhang, "Major 5G Waveform Candidates: Overview and Comparison," \textit{Signal Processing for 5G: Algorithms and Implementations}, IEEE, 2016, pp.169-188.

\end{thebibliography}
\end{document}